\documentclass{aa}

\usepackage{ulem}
\usepackage{graphicx}
\usepackage{amsmath}
\usepackage{subfigure}
\usepackage[longnamesfirst]{natbib}
\usepackage{txfonts}
\bibpunct{(}{)}{;}{a}{}{,}

\headnote{Research Note}
\title{What are the temperatures of T\,Tauri stars?}
\subtitle{Constraints from coeval formation of young eclipsing binaries\thanks{The appendices A and B are only available in electronic form}}

\author{M. Ammler\inst{1}, V. Joergens\inst{2}, R. Neuh\"auser\inst{1}}

\institute{AIU Jena, Schillerg\"a{\ss}chen 2-3, D-07745 Jena, Germany
\and Sterrewacht Leiden/ Leiden Observatory, Niels Bohrweg 2, NL-2333 CA Leiden, The Netherlands}

\offprints{M. Ammler,
\email{ammler@astro.uni-jena.de}
}

\date{Received / Accepted}

\shortcites{Perrin98}
\shortcites{Basri00}
\shortcites{Baraffe02}
\shortcites{Blackwell91}
\shortcites{Burgasser02}
\shortcites{Burgasser01}
\shortcites{Casey98}
\shortcites{CBAH00}
\shortcites{Code76}
\shortcites{Covino00}
\shortcites{Covino01}
\shortcites{Covino04}
\shortcites{Guilloteau99}
\shortcites{Hauschildt99a}
\shortcites{Hauschildt99b}
\shortcites{Henry94}
\shortcites{Ridgway80}
\shortcites{Leggett96}
\shortcites{Leggett98}
\shortcites{Leggett00}
\shortcites{BCAH98}
\shortcites{Richichi98}
\shortcites{vanBelle99}
\shortcites{Jones95}
\shortcites{Fuhrmann97}
\shortcites{Luhman03}
\shortcites{Montalban04}
\shortcites{SDF00}
\shortcites{Segransan03}
\shortcites{Stassun04}
\shortcites{Steffen01}
\shortcites{White99}
\shortcites{Yi03}

\defcitealias{BCAH98}{BCAH98}
\defcitealias{PS99}{PS99}

\abstract{We show how the assumption of coeval formation can be used to constrain the effective temperatures of the components of young eclipsing double-lined spectroscopic binaries. Our method extends the approach of \citet{White99} to a two-step analysis. The first step compares evolutionary models to the observed masses and radii and selects those models that predict ages that are consistent with coeval formation. The second step then uses these models to constrain the effective temperatures. We applied the method on literature values of the stellar parameters of the eclipsing binaries \object{RX\,J0529.4+0041\,A} and \object{V1174\,Ori} and confirm that \object{V1174\,Ori~A} has dwarf-like temperatures at an age of 9\,Myrs, while we cannot draw any conclusions for \object{RX\,J0529.4+0041\,A} and \object{V1174\,Ori~B}.
Considering these binaries, we find that none of the evolutionary models gives coeval solutions simultaneously in mass, radius and effective temperature.

}

\begin{document}

\titlerunning{What are the effective temperatures of T\,Tauri stars?}
\authorrunning{M. Ammler et al.}

\maketitle

\section{Introduction}



Effective temperatures can be determined directly by measuring the star's angular diameter and its emerging total flux. While this works well for giant stars and very nearby main-sequence stars, the resolution of new interferometers now approaches also the red dwarf regime \citep{Segransan03}. Furthermore, effective temperatures can be derived with semi-direct methods, e.g. the infrared flux method which relies partly on stellar atmosphere modelling \citep[see][]{Blackwell91}. \citet{Fuhrmann04a} determined precise effective temperatures of several hundred nearby stars by modelling the Balmer line wings. The application of these methods to T\,Tauri stars (TTSs) is difficult. The main problem is the relatively large distance to nearby star forming regions requiring very high angular resolution to measure diameters directly. Furthermore, in the case of classical TTSs, measurements are difficult because of the  UV and IR emission from the accretion disk. Moreover, activity of TTSs results in strong spectral features which hamper the spectroscopic methods. Therefore, effective temperatures of TTSs usually have to be estimated by means of temperature calibrations. 

In a few cases, the light curves of pre-main sequence (PMS) eclipsing double-lined spectroscopic binaries (SB2s) allow the direct determination of the ratio of the components' effective temperatures. At first there is no information on the individual temperatures of the components which have to be constrained by external information. The type of the external information generally depends on the binary studied and the available observational data. One common approach is the derivation of the primary temperature from its spectral type or colour index. As an example,
\citet{Covino04} estimated the effective temperature of \object{RX\,J0529.4+0041\,Aa} from its spectral type by means of calibrations. The assumption of temperature scales for dwarf stars is supported by the surface gravities of the components \citep{Covino05}. Furthermore, they created a synthetic composite spectrum of \object{RX\,J0529.4+0041\,A} which is based on the derived stellar parameters and the models of \citet{Hauschildt99a,Hauschildt99b}. This spectrum was found to be consistent with the combined UBVRIJHK fluxes.

The derivation of the effective temperature from colour index or spectral type depends on the luminosity class, which is a priori unknown for PMS stars. Although the surface gravity can be precisely determined for PMS eclipsing SB2s, we think that it remains unclear whether main-sequence or post main-sequence temperature scales are applicable to PMS stars. Therefore, we suggest a further constraint to be applied for PMS stars which follows \citet{White99}, and is based on the assumption of coevality and the use of evolutionary models.

Their approach is the only method which determines effective temperatures of PMS stars independently from any assumptions on the luminosity class. They used the assumption of coeval components of the quadruple \object{GG\,Tau} to constrain their temperatures and find intermediate temperatures between giants and dwarfs. The method was applied by \citet{Luhman99} and \citet{Luhman03} to create a temperature scale for intermediate luminosity classes at spectral types M for use with the evolutionary models of \citet{BCAH98} and \citet{CBAH00}. 

We extend the approach of \citet{White99} in Sect.~\ref{sect:method} for PMS eclipsing SB2s and apply it to \object{RX\,J0529.4+00.41\,A} and \object{V1174\,Ori} in Sect.~\ref{sect:application}.


\section{A new method -- extending the approach of \citet{White99}}
\label{sect:method}
\citet{White99} obtained the temperatures of the \object{GG\,Tau} components by comparing their luminosities to evolutionary models in the HR diagram. The assumption of coevality requires the components to be located on the same isochrones and therefore constrains their effective temperatures (see Sect.~\ref{sect:luhman99i} for further details). Although the temperatures are allowed to vary within a relatively large range between dwarf and giant values, coeval solutions are only found for the \citet{BCAH98} \citepalias{BCAH98} isochrones. 
This coevality test is not independent from the derivation of the temperatures of the components. An independent test of the models (e.g. with individual masses and radii) cannot be done with GG\,Tau. In contrast to \object{GG\,Tau}, individual masses and radii of the components of eclipsing SB2s are well determined. Before constraining the effective temperatures, we can therefore first rule out all evolutionary models that do not give coeval solutions for the measured masses and radii. We retain only the models with coeval solutions supporting the accuracy of the resulting effective temperatures.

In our method we first compare the fundamental masses and radii of the binary components to the predictions of evolutionary models by \citet{DM94}, \citet{DM97}, \citetalias{BCAH98}, \citet{PS99} \citepalias{PS99}, \citet{SDF00} and \citet{Yi03} (see Table~\ref{tab:models}). Then, we keep only the models which give coeval solutions for masses and radii. As in \citet{White99}, the adopted temperatures may vary within a large range spanned by the various dwarf and giant scales as presented below. We obtain temperatures from each of these scales for the primaries and calculate the secondary temperatures from the temperature ratio given by the light curve analyses of \citet{Covino04} and \citet{Stassun04}. Finally, we compare these temperatures to the selected evolutionary models in the $T_\mathrm{eff}-R$ diagram to find consistent temperatures.

\begin{table}
\caption{Overview of the used evolutionary models from \citet{DM94} (DM94), \citet{DM97} (DM97), \citetalias{BCAH98}, \citetalias{PS99}, \citet{SDF00} (SDF00) and \citet{Yi03} (YKD03) with the choices of free parameters. The table is restricted to those parameters which are necessary to distinguish between the models.}
\label{tab:models}
\centering
\begin{tabular}{l|llll}
\hline
\hline
model&metallicity&Y&$X_D/10^{-5}$&convection\\
\hline
DM94  &                &          &                  &MLT\\
              &                &          &                  &CM\\
\hline
DM97  & $Z=0.01$       & 0.26     & $2\times10^{-5}$ & \\
              & $Z=0.01$       & 0.26     & $4\times10^{-5}$ & \\
              & $Z=0.01$       & 0.28     & $1\times10^{-5}$ & \\
              & $Z=0.01$       & 0.28     & $2\times10^{-5}$ & \\
              & $Z=0.01$       & 0.28     & $4\times10^{-5}$ & \\
\hline
BCAH98& $[M/H]=0.0$    & 0.275    &                  &MLT ($\alpha=1.0$)\\
              & $[M/H]=0.0$    & 0.275    &                  &MLT ($\alpha=1.5$)\\
              & $[M/H]=0.0$    & 0.282    &                  &MLT ($\alpha=1.9$)\\
              &$[M/H]=-0.5$    & 0.250    &                  &MLT ($\alpha=1.0$)\\
\hline
PS99  &                & 0.28     &                  &MLT ($\alpha=1.5$)\\
\hline
SDF00 & $Z=0.01$       & 0.256    &                  &\\
                          & $Z=0.02$       & 0.277    &                  &\\
                          & $Z=0.02$       & 0.277    &                  &$+$overshooting\\
                          & $Z=0.03$       & 0.297    &                  &\\
                          & $Z=0.04$       & 0.318    &                  &\\
\hline
YKD03  & $Z=0.01$       &          &                  &\\
                      & $Z=0.023$      &          &                  &\\
                      & $Z=0.03$       &          &                  &\\
\hline
\end{tabular}
\end{table}

Compared to \citet{White99}, we apply a larger number of temperature scales. We compiled 14 effective temperature scales for several luminosity classes, focussing on spectral types G0-M9. We consider these scales to be representative for the large amount of temperature scales in the literature, and we believe that they reflect the range of physically possible temperatures at a certain spectral type.

\begin{figure}
\resizebox{\hsize}{!}{\includegraphics{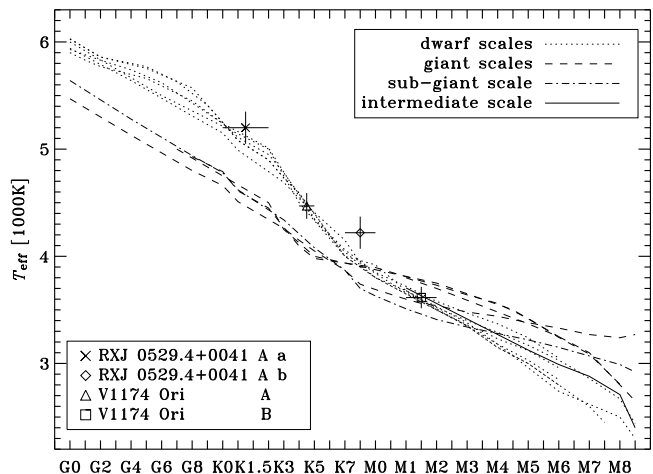}}
\caption{\label{fig:scales_ecl_bin}Overall view of the adopted temperature scales in the spectral type vs. effective temperature diagram. Additionally, the measured spectral types and effective temperatures of the eclipsing binaries are shown (from Table~\ref{tab:EclBin}). The error bar of \object{V1174\,Ori}\,B is only a very rough estimate.}
\end{figure}

Fig.~\ref{fig:scales_ecl_bin} displays an overview of these scales in a spectral type vs. effective temperature diagram. References and general information are listed in Table~\ref{tab:scales}. Temperature values at each spectral type are tabulated for spectral types G0-M9 in App.~\ref{app:tables} (Table~\ref{tab:dwarfs} for dwarf scales; Table~\ref{tab:giants} for other scales). Furthermore, we compiled or derived intrinsic error estimates for the temperature scales (Table~\ref{tab:scales} and App.~\ref{app:errors}).
Fig.~\ref{fig:scales_ecl_bin} shows that the temperature scales for stars of spectral types earlier than K7 are divided into two distinct bundles, one for the giants and one for the dwarfs, with giant temperatures being up to $\sim500\,$K cooler than the dwarf temperatures. The spread in dwarf temperatures is up to $300\,$K for same spectral type, and for giant temperatures it is $\sim200\,$K, giving a total span of physically possible effective temperatures at the same spectral types of up to $800\,$K.
At the spectral type $\sim$K7, we find a crossing point where all scales provide similar temperatures. At later spectral types the giant temperatures are generally hotter than the dwarf temperatures. The overall temperature spread increases again and peaks at $1000\,$K for the very late spectral types.

\begin{table*}
\caption{Adopted temperature scales with validity range in MK spectral type, luminosity class, method of construction, intrinsic errors and validity range of the intrinsic errors. 
We distinguish between calibrations which are based directly on effective temperature measurements and such that are only derived from calibrations of other authors. We refer to the former as to {\it primary} and to the latter as to {\it derived} scales. Scales which are based on older ones but account for new measurements are indicated by `improved'. The origin of the intrinsic errors is explained in App.~\ref{app:errors}. 
}
\label{tab:scales}
\centering
\begin{tabular}{l|lllll}
\hline
\hline
reference&sp. type&lum. class&construction&intrinsic errors&validity range of int. errors\\
\hline
\citet[table~2]{Bessell79}$^1$&B7-M6&dwarfs&primary&220\,K$^2$&earlier G2\\
&&&&$\Delta{\log}T_\mathrm{eff}=0.035^2$&later than G2\\
&&&&&and earlier than K7\\
\citet[table~2]{Bessell91}$^1$& K7-M7.5&dwarfs&primary&290\,K$^2$&whole range\\
\hline
\citet[table~3]{Bessell79}&G7-M6&giants&primary&220\,K$^2$&whole range\\
\hline
\citet[table~7]{CK79}&O9.5-M6&dwarfs&derived&300\,K$^2$&whole range\\
\hline
\citet[table~5]{dJN87}&O3-M9&dwarfs&primary&$\Delta{\log}T_\mathrm{eff}=0.021^3$&whole range\\
\hline
\citet[table~5]{dJN87}&O3-M9&subgiants&primary&$\Delta{\log}T_\mathrm{eff}=0.021^3$&whole range\\
\hline
\citet[table~5]{dJN87}&O3-M9&giants&primary&$\Delta{\log}T_\mathrm{eff}=0.021^3$&whole range\\
\hline
\citet[table~4]{HSS94}&F0-M6&dwarfs&derived&$\Delta{\log}T_\mathrm{eff}=0.015^2$&earlier than K7\\
&&&&290\,K$^2$&later than K7\\
\hline
\citet[table~5]{KH95}&B0-M6&dwarfs&improved&$\Delta{\log}T_\mathrm{eff}=0.015^2$&whole range\\
\hline
\citet[table~5]{Perrin98}&G8-M8&giants&primary&220\,K$^2$&earlier than M6\\
&&&&79\,K$^3$&M6\\
&&&&94\,K$^3$&M7\\
&&&&42\,K$^3$&M8\\
\hline
\citet[table~2]{Luhman99}&M1-M9&giants&derived&270\,K$^2$&earlier than M7\\
&&&&100\,K$^2$&M7\\
&&&&50\,K$^2$&M8\\
&&&&150\,K$^2$&M9\\
\hline
\citet[figure~8]{Luhman03}&M1-M9&dwarfs&improved&80\,K$^2$&whole range\\
\hline
\citet[table~8]{Luhman03}&M1-M9&intermediate&improved&80\,K$^2$&earlier than M6.5\\
&&&&100\,K$^2$&later than M6.5\\
\hline
\citet[table~7.6]{Tokunaga00}&O9-M6&dwarfs&derived&100\,K$^3$&later spectral types\\
\hline
\end{tabular}
\begin{flushleft}
$^1$We only used a combination, i.e. \citet{Bessell79} for spectral types earlier than K7 and \citet{Bessell91} for spectral types later than K7.\\
$^2$Intrinsic errors were derived by us or taken from another work.\\
$^3$Intrinsic errors were adopted from the original work.
\end{flushleft} 
\end{table*}



\section{Application to known eclipsing binaries}
\label{sect:application}
\subsection{\object{RX\,J0529.4+0041\,A} and \object{V1174\,Ori}}
\label{sect:eclbin}
In the past years, several eclipsing SB2s with low-mass PMS components have been found: \object{RX\,J0529.4+0041\,A} \citep{Covino04}, \object{V1174\,Ori} \citep{Stassun04}, \object{TY\,CrA} \citep{Casey98} and \object{EK\,Cep} \citep{HE84,Popper87}. A further system with known masses, the astrometric binary \object{NTT\,045251+3016}, was found by \citet{Steffen01}. We use \object{RX\,J0529.4+0041\,A} and \object{V1174\,Ori} (see Table~\ref{tab:EclBin} for the adopted physical parameters) for the further analysis. We do not consider \object{TY\,CrA} and \object{EK\,Cep} because their higher-mass primaries are already on the main sequence and therefore, coevality cannot be tested with the PMS evolutionary models\footnote{However, such high-mass main-sequence primaries enable the direct use of main-sequence temperature scales to determine the temperature of the lower-mass PMS secondary!}. Furthermore, we did not take into account \object{NTT\,045251+3016} because its radii cannot be measured directly.


Similar to \citet{Covino04}, \cite{Stassun04} determined the effective temperature of the primary on grounds of its spectral type and the scale of \citet{SK82} while the temperature of the secondary was constrained by the temperature ratio from the light curve analysis.

\begin{table}
\caption{The physical parameters of the two young eclipsing binaries \object{RX\,J0529.4+0041\,A} and \object{V1174\,Ori} as adopted from \citet{Covino04} and \citet{Stassun04}, respectively. We use their temperature solutions but also determine a set of additional effective temperatures using other temperature scales (see Sec.~\ref{sect:temp}).}
\label{tab:EclBin}
\centering
\begin{tabular}{l|ll}
\hline
\hline
&\object{RX\,J0529.4+0041 Aa}&\object{RX\,J0529.4+0041 Ab}\\
\hline
$M\,[M_{\odot}]$&$1.27\pm0.01$&$0.93\pm0.01$\\
$R\,[R_{\odot}]$&$1.44\pm0.05$&$1.35\pm0.05$\\
${\log}g\,[cgs]$&$4.22\pm0.02$&$4.14\pm0.02$\\
$T_\mathrm{eff}\,$[K]&$5200\pm150$&$4220\pm150$\\
spectral type&K1$\pm$1&K7-M0\\
\hline
&\object{V~1174\,Ori\,A}&\object{V~1174\,Ori\,B}\\
\hline
$M\,[M_{\odot}]$&$1.009\pm0.015$&$0.731\pm0.008$\\
$R\,[R_{\odot}]$&$1.339\pm0.015$&$1.065\pm0.011$\\
${\log}g\,[cgs]$&$4.19\pm0.01$&$4.25\pm0.01$\\
$T_\mathrm{eff}\,$&$4470\pm120$&$3615\pm100$\\
spectral type&K4.5$\pm$0.5&M1.5\\
\hline
\end{tabular}
\end{table}

While Fig.~\ref{fig:scales_ecl_bin} shows that indeed the temperatures are consistent with the dwarf relations, it is puzzling that \object{RX\,J0529.4+0041\,A}b is somewhat hotter than both dwarf and giant temperatures. It is important to keep in mind here that these individual temperatures rely on much weaker constraints than the temperature ratio which is well known for both binaries from the eclipse light curves. 


\subsection{First step -- selecting appropriate evolutionary models}
\label{sect:test}

\begin{figure*}
\begin{center}
\subfigure{\includegraphics[width=8.5cm]{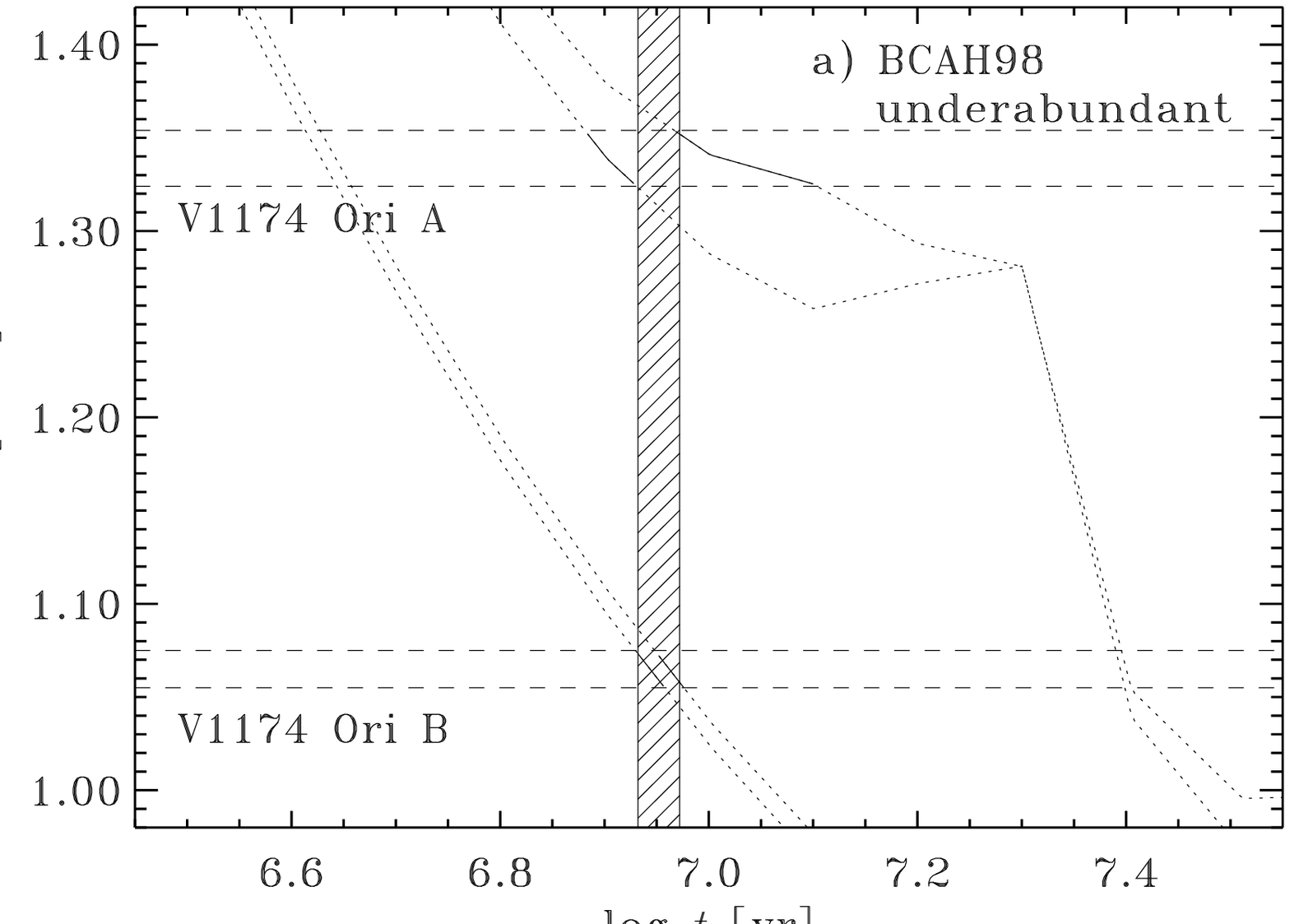}}
\subfigure{\includegraphics[width=8.5cm]{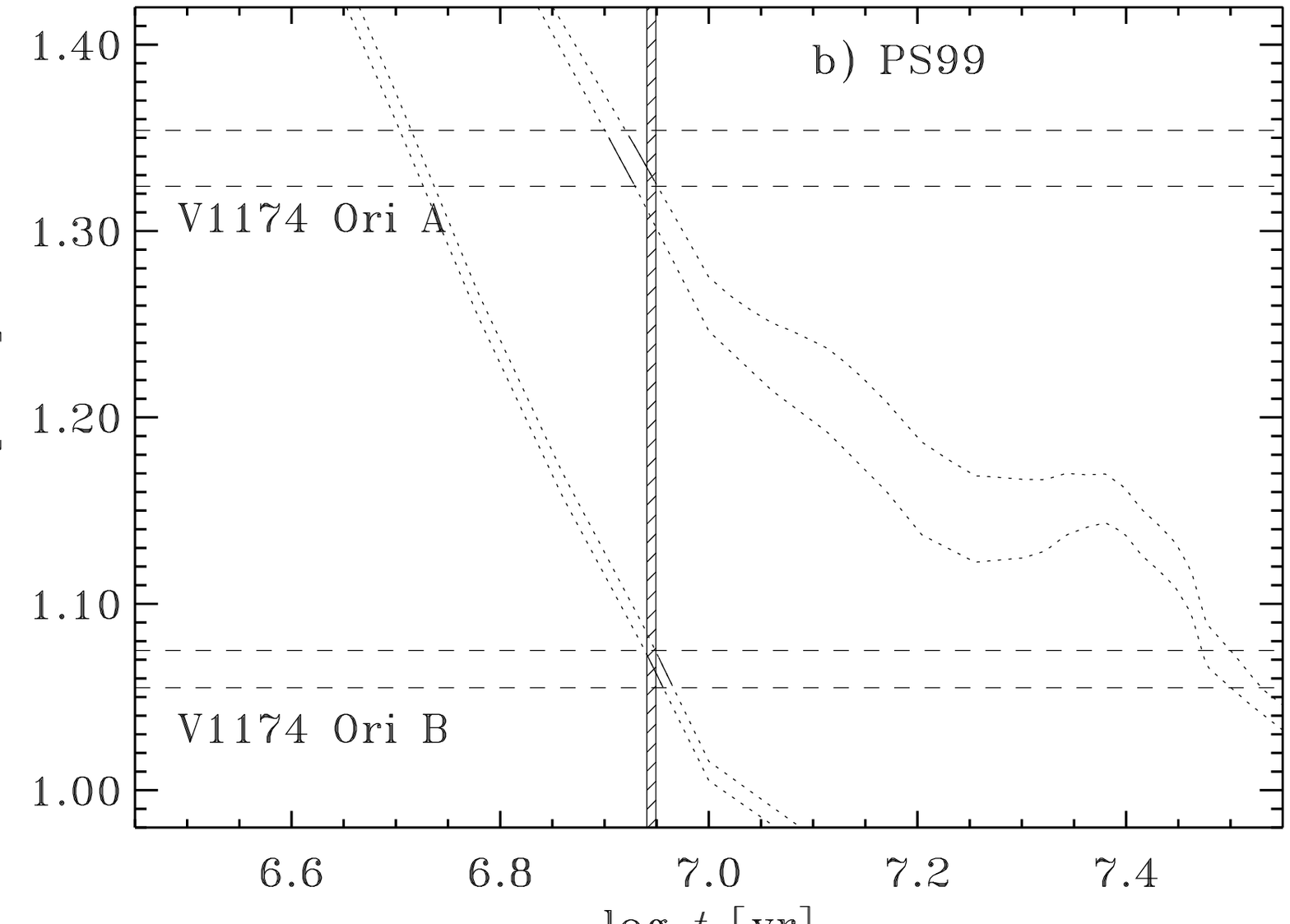}}
\end{center}
\caption{\label{fig:v11_MR}The theoretical evolution of the radii of \object{V1174\,Ori}~A \& B is compared to the observed values. We only show the cases with coeval solutions: {\bf a)} \protect\citetalias[][underabundant]{BCAH98} and {\bf b)} \protect\citetalias{PS99}. The dashed horizontal lines indicate the upper and lower limits of the radius measurements. Evolutionary tracks for the upper and lower limits of the dynamical mass measurements were interpolated in the indicated models (dots). Solid lines represent the parts of the interpolated tracks which are consistent with the observed radii. These span the possible age of the individual binary components. The hatched region represents the coeval solution of $\sim9\,$Myrs.}
\end{figure*}

For the case of \object{V1174\,Ori}, Fig.~\ref{fig:v11_MR} illustrates how we found evolutionary models giving coeval solutions for the radii. If a coeval solution can be found, we conclude that the specific set of models can be used to constrain the effective temperatures.

No single set of evolutionary models provides a coeval solution for the masses and radii of \object{RX\,J0529.4+0041\,A}a \& Ab, so we do not consider this system any further. In the case of \object{V1174\,Ori}, we find that only the metal-poor models of \citetalias{BCAH98} and the models of \citetalias{PS99} are consistent with coeval formation (Fig.~\ref{fig:v11_MR}). The corresponding age is $\sim9\,$Myrs. In contradiction to \citet[][fig.~19]{Stassun04}, we do not find coeval solutions when using the models of \citet{SDF00}\footnote{Their fig.~19 suggests coevality if inspected by eye. However, using a more quantitative analysis by interpolating isochrones with small time steps, we find that coevality no longer holds.}.
Concerning \object{RX\,J0529.4+0041\,A}b, we point out that \citet{Covino04} found inconsistencies between its rotational velocity and its radius. The synchronisation condition would indicate the measured radius is overestimated by about 20\,\%, possibly reconciling the age discrepancy with \object{RX\,J0529.4+0041\,A}a. Nevertheless, we give higher weight to the results of the light curve analysis at hand.

\subsection{Second step -- constraining effective temperatures}
\label{sect:temp}
\begin{figure*}
\begin{center}
\subfigure{\includegraphics[width=8.5cm]{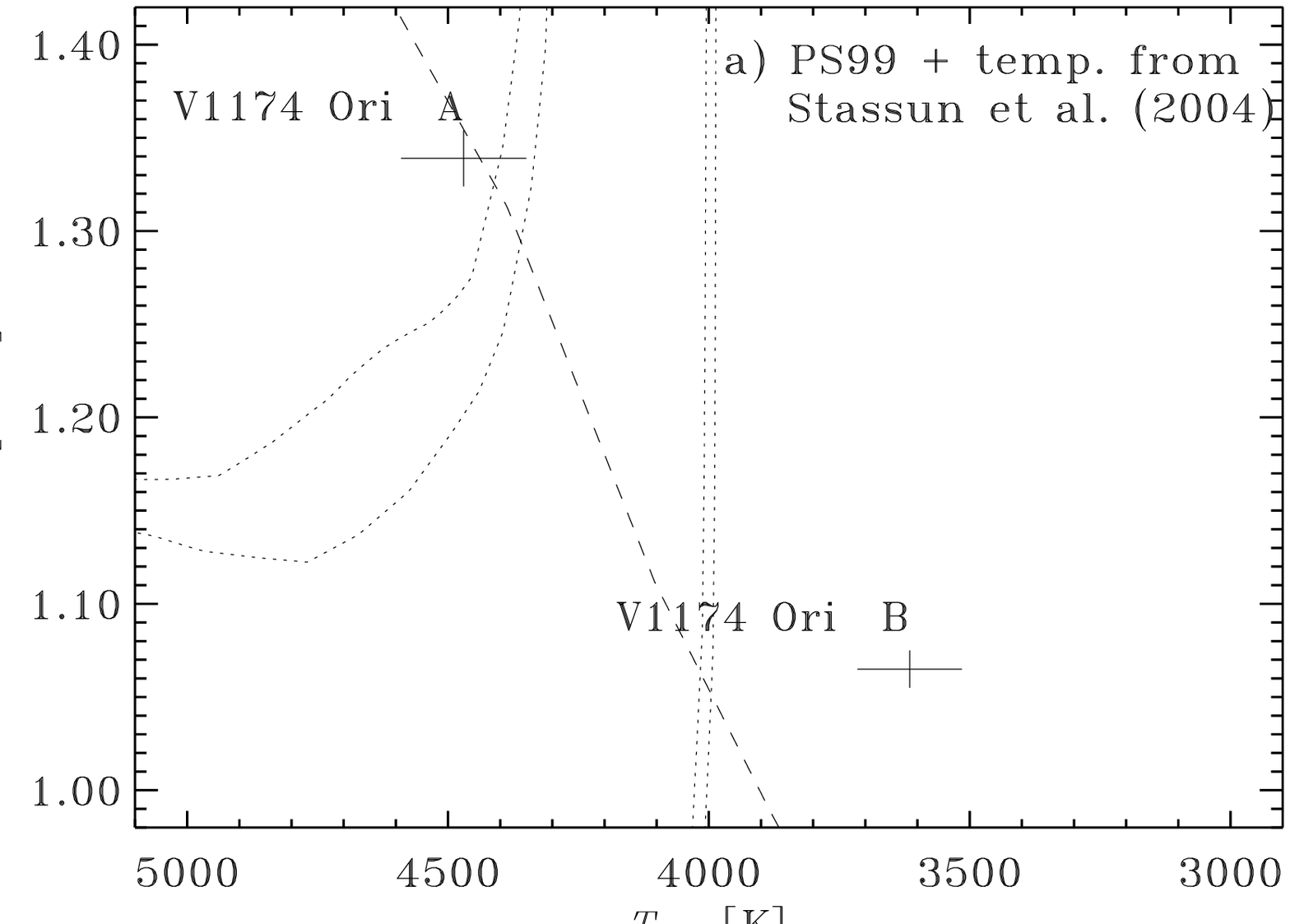}}
\subfigure{\includegraphics[width=8.5cm]{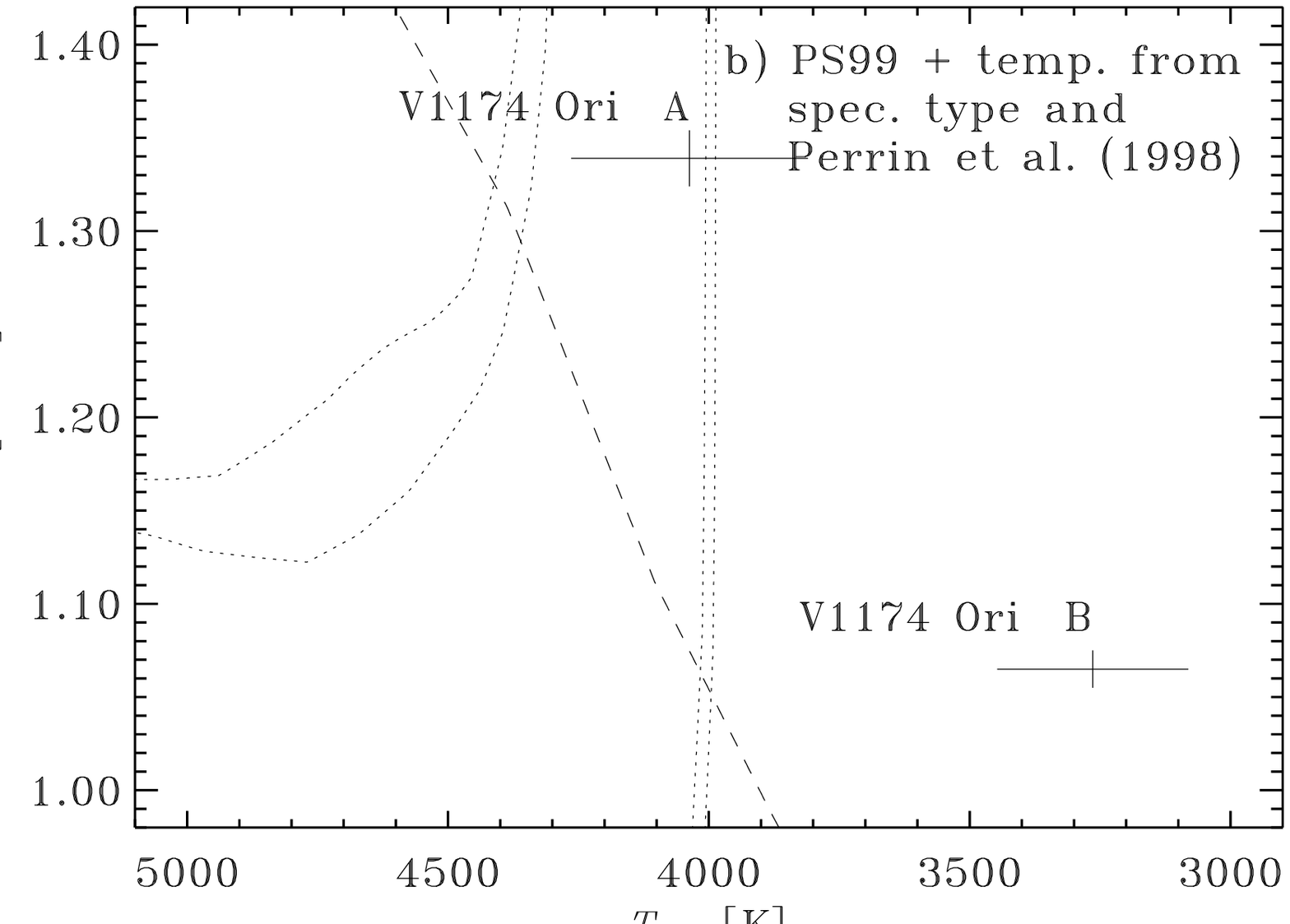}}
\end{center}
\caption{\label{fig:v11_Kiel}$T_\mathrm{eff}-R$ diagrams with the error bars of \object{V1174\,Ori\,A} \& B. Evolutionary tracks for the upper and lower limits of the dynamical mass measurements (Table~\ref{tab:EclBin}) (dots) were interpolated in the indicated tracks. The dashed line shows the theoretical isochrone at the common age of 9\,Myrs found in Sect.~\ref{sect:test}. {\bf a)} In this case, the temperatures were adopted from Table~\ref{tab:EclBin}. The effective temperature of \object{V1174\,Ori\,A} is consistent with the corresponding theoretical tracks whereas the temperature of \object{V1174\,Ori\,B} is not. {\bf b)} Temperatures were derived with the giant scale of \protect\citet{Perrin98} and are inconsistent with the models.}
\end{figure*}

We now reconsider the effective temperatures of \object{V1174\,Ori\,A} \& B, the only system where evolutionary models provide coeval solutions in the previous section. We interpolate the primary's spectral type in each conversion table (Tables~\ref{tab:dwarfs} and \ref{tab:giants}) taking into account intrinsic errors of the particular scale (Table~\ref{tab:scales}) and errors of the measured spectral types (Table~\ref{tab:EclBin}). 
The spectral type of \object{V1174\,Ori\,A} is earlier than M0 so that the scales of \citet{Luhman99} and \citet{Luhman03} do not apply. The secondary's temperatures were calculated from the primary's temperatures and the measured temperature ratio. In order to compare with the predictions of the evolutionary models, we use the $T_\mathrm{eff}-R$ diagram instead of the HR diagram because the radii of \object{V1174\,Ori\,A} \& B are much better constrained than the luminosities, which are subject to systematic uncertainties such as extinction and distance. Fig.~\ref{fig:v11_Kiel} illustrates how we found temperatures which are consistent with the appropriate evolutionary tracks of \citetalias{BCAH98} and \citetalias{PS99}. 

These models suggest temperatures of $\sim4500\,$K \citepalias{BCAH98} and $\sim4400\,$K \citepalias{PS99}, respectively, for \object{V1174\,Ori\,A}. That means they rule out giant-like temperatures (cf. Fig. \ref{fig:v11_Kiel}b) but do not distinguish between almost all the other scales. The subgiant scale of \citet{dJN87} yields temperatures which are inconsistent with the underabundant models of \citetalias{BCAH98} but consistent with the models of \citetalias{PS99}. Also the temperature of \object{V1174\,Ori\,A} from \citet{Stassun04} is consistent with both models.

In the case of \object{V1174\,Ori\,B} however, the two models do not agree with any temperature which we derived, not even with the temperature from Table~\ref{tab:EclBin}. The predicted temperatures are significantly hotter (by a few hundred K) than any of the derived dwarf or giant temperatures. Furthermore, coevality is no longer fulfilled in the $T_\mathrm{eff}-R$ diagram. 
A similar effect was found by \citet{HW04} who systematically compared evolutionary models to PMS and main-sequence stars with known masses but not yet including \object{V1174\,Ori}. Our results are mostly in line with the findings of \citet{Stassun04} who compared all PMS stars with observationally determined masses to evolutionary models in both the HR diagram and the more fundamental mass-radius diagram. 

\subsection{Discussion}
\label{sect:disc}
Our test in the current set-up still suffers from some shortcomings. First of all, we have only used $1\,\sigma$ errors. Furthermore, we required strict coevality although the evolutionary models describe single stars and not close binaries. We do not know to what degree coevality is realised in real binaries. The formation of both components might have set in at slightly different times. Furthermore, magnetic fields and rotation are known to slow down star formation but are not taken into account in classical hydrostatic models, possibly causing some additional real or apparent relative age difference. Moreover, the hydrostatic models do not take into account the dynamical phases before the quasi-hydrostatic PMS contraction. Dynamical calculations by \cite{WT03} predict differences in the contraction behaviour on the PMS at different stellar masses. In addition, theoretical luminosities and temperatures are based on stellar atmosphere models which still have to cope with several problems especially for young stars. Following \citet{Wuchterl01}, \citet{Baraffe02} and \citet{WT03}, one should not rely on classical hydrostatic models for ages of a few Myrs. 

Moreover, a limited number of stellar parameters have been determined by observations for the systems considered here. Further observational constraints, e.g. on metallicities, should be included in order to find matching models. For example, it seems rather doubtful that of all the \citetalias{BCAH98} models, only the metal-poor model gives coeval radius solutions for \object{V1174\,Ori}. Of course, these shortcomings also apply to all such comparisons which have been performed by other authors.


\section{Summary and conclusions}
Up to now, precise temperature constraints for PMS stars are only available from PMS eclipsing SB2s. While the effective temperature ratio can be precisely determined from the light curves, individual temperatures of each component are based on more uncertain external constraints. Therefore, we propose in this paper an extension of the method of \citet{White99} to constrain effective temperatures of the components of PMS eclipsing SB2s. The application of our method is at the moment restricted to the PMS eclipsing SB2s \object{RX\,J0529.4+0041\,A} and \object{V1174\,Ori}.

The method first compares the empirical masses and radii of the binary components to the predictions of evolutionary models and selects those models which give coeval solutions. Secondly, a set of temperatures is derived from the primary's spectral type using several temperature scales. Then, temperatures for the secondary are calculated from each primary temperature and the temperature ratio, which is known from the light curve analysis. Finally, these temperatures are compared to the predictions of the selected models in the $T_\mathrm{eff}-R$ diagram. 

For the second step of the procedure, we use our compilation of temperature scales for dwarfs, giants and intermediate luminosity classes from the literature. If not yet available, we derived intrinsic uncertainties of the scales. The scales are different due to the different derivation methods. Even scales for the same luminosity class differ by up to 300\,K at spectral types G and up to 1000\,K at late-M types. Intrinsic uncertainties in the individual temperature scales typically amount to a few hundred K.

Applying the first step of our analysis to the PMS binaries, we find that none of the models is consistent with coevality of \object{RX\,J0529.4+0041\,A}a \& Ab while coevality of \object{V1174\,Ori}~A \& B is only consistent with the \citet{PS99} and the metal-poor \citet{BCAH98} models. However, it seems doubtful that V1174\,Ori is that metal-poor. 

In the second step, neither the dwarf nor the giant temperatures nor the temperatures estimated by \citet{Stassun04} are consistent with the models in the case of \object{V1174\,Ori~B}. \object{V1174\,Ori\,A} alone gives strong evidence for dwarf-like temperatures at spectral types mid-K and ages of $\ge9\,$Myrs corroborating the use of dwarf scales for \object{RX\,J0529.4+0041\,A} by \citet{Covino04} and for \object{V1174\,Ori} by \citet{Stassun04}.

As in the studies by \citet{Luhman99} and \citet{Luhman03}, our method may provide a new intermediate temperature scale for young stars once enough accurate stellar parameters from PMS eclipsing binaries are available.


\begin{acknowledgements}
We are grateful for the useful suggestions by G. Wuchterl and want to thank K. Fuhrmann and T. Gehren for the interesting discussions on effective temperature determination. We gratefully acknowledge linguistic improvements on the paper by G. Torres. We thank E. Covino, Th. Schmidt-Kaler and S. Kenyon for giving useful remarks. We further thank F. Comer{\'o}n, G. Herbig, and A. Richichi for reading the paper and giving interesting suggestions. We are further thankful for useful comments by H.-G. Ludwig. MA acknowledges financial support from the episcopal study foundation `Cusanuswerk'. VJ acknowledges support from the Deutsche Forschungsgemeinschaft (Schwerpunktprogramm `Physics of star formation') and from the European community by a Marie Curie Individual Fellowship. MA, VJ and RN did part of this work when they were together at MPE Garching in 2001/2002. 
\end{acknowledgements}

\bibliographystyle{aa}
\bibliography{/astro/ammler/publications/ref}
\Online
\appendix

\section{Derivation of intrinsic scale errors}
\label{app:errors}
\subsection{\citet{Bessell79,Bessell91}}
\citet{Bessell79} created temperature scales for dwarfs and giants (their tables 2 and 3, resp.). The scale for M dwarfs has been improved by \citet{Bessell91} so that we consider only the combined dwarf scale (see Table \ref{tab:scales}). For spectral types earlier than K7 the dwarf scale of \citet[][table~2]{Bessell79} is applied and for later types the work of \citet{Bessell91}. 

\subsection{\citet{Bessell79}}
\label{sect:bessell79}
\citet{Bessell79} does not give any errors for the temperature scales. In order to get an estimation for these errors, we analyse the origin of the temperatures in more detail. 

The dwarf temperature scale is based upon a fit-by-eye to the temperatures of \citet{Code76} \citep[cf.][table~1, figure~10]{Bessell79} for early spectral types.
The independent variable in the fit is not the spectral type but the Str\"omgren index ($b-y$) which has to be converted to spectral type. The temperatures derived by \citet{Code76} and the fit in \citet{Bessell79} allow us to calculate the standard deviation. This value of $\Delta{\log}T=0.035\,$dex now represents an estimate for the uncertainties of the scale for early spectral types. We did not take into account any errors that might be introduced by the conversion from ($b-y$) to spectral type.

The temperature scale for dwarfs for temperatures between $4000\,$K and the temperature of the sun is based on the relation between $(V-I)$ and the temperatures of giants.

The temperatures for cooler dwarfs are taken from the work of \citet{Veeder74} who fitted black body distributions to spectral energy distributions between $0.4\,{\rm \mu}$m and $3.5\,{\rm \mu}$m. Those temperatures are not used in our work because they have been superseded by \citet{Bessell91}.

The temperatures for giants cooler than the sun were adopted from the scale of \citet{Ridgway80}. 
In order to derive their temperature scale, \citet{Ridgway80} only used direct effective temperature measurements with errors lower than $\pm250\,$K (cf. their table 3) and performed a fit-by-eye in the ($T_{\rm eff}-T_\mathrm{C}$) diagram. The colour temperatures $T_{\rm C}$ resulted from a fit of black body distributions to the continuum flux at $8500\,${\AA} and $10500\,${\AA}. The colour temperatures have been converted into spectral types \citep{WY79} in order to obtain a relation between spectral type and effective temperature. The work of \citet{Ridgway80} does not provide any uncertainties for the temperature scale. In order to get an idea of the uncertainty of the scale in \citet{Bessell79}, we calculated the standard deviation of the fit-by-eye and obtained $220\,$K. Approximating the fit-by-eye with a second degree polynomial yields a similar result, when excluding the problematic stars BS~5301, HD~75156 and HD~29051. We did not take into account any errors which might be introduced by the conversion from $T_\mathrm{C}$ to spectral type.

\subsection{\citet{Bessell91}}
\label{sect:bessell91}



\citet{Bessell91} combined in their figure~11 their own colour measurements and effective temperatures from the literature in the $(R-I)$-$T_{\rm eff}$ diagram for early-type stars and in the $(I-K)$-$T_{\rm eff}$ diagram for late-type stars. The fundamental temperatures of YY~Gem and CM~Dra were included. The data were approximated by third order polynomials.
The fundamental temperatures obviously have significantly higher weight. The origin of the spectral types in their table~2 was not clarified in \citet{Bessell91}.

\citet{Bessell91} do not provide any errors for the temperature scale. In order to get an estimate for the uncertainties implied in the application of this scale, we calculated the standard deviation of the polynomial ($R-I$) calibration for stars in their figure~11\,a. It amounts to $290\,$K. We used only some 20 stars for which we could reproduce the required data following the information given in \citet{Bessell91}. We adopt the calculated standard deviation also for the later spectral types as is justified by a short inspection by eye of their figure~11\,b. 

\subsubsection{Summary: the intrinsic errors}
The application of the giant scale implies an error of $220\,$K at all spectral types. The error for the combined dwarf temperature scale is approximately $\Delta{\log}T_\mathrm{eff}=0.035\,$dex for spectral types earlier than that of the sun, $\pm220\,$K for later types earlier than K7 and $\pm290\,$K for the latest types. We consider both scales to be primary scales because they were constructed directly with temperatures of individual stars.


\subsection{\citet{CK79}}
\citet{CK79} 
created a temperature scale (their table~7) which was recommended by \citet{SK82} for the analysis of pre-main sequence stars. It is based on scales from the literature. Though \citet{CK79} found that the luminosity class of T Tauri stars is between that of giants and dwarfs, they use a temperature scale which is based on dwarf scales because those were available more easily. Thus, to be consistent, we consider this temperature scale to be a dwarf scale. According to \citet{CK79} the adoption of this scale for T\,Tauri stars implies an overestimation of the temperatures by 5\%. We find that the temperatures would be overestimated only for spectral types earlier than M if we would assign giant-like properties to T\,Tauri stars.


The temperature scale of \citet{CK79} is based on the work of \citet{Vardya70} who applied temperatures of \citet{Johnson65} and \citet{Johnson66}, inferred from observed apparent diameters and bolometric fluxes. The temperatures for the spectral types K5 and K7 are obviously adopted from \citet{Johnson66} and for M from \citet{Vardya70}. Temperatures for fractional spectral types and for K7-M0 are apparently found by interpolation.

\citet{CK79} do not provide estimates for the uncertainties of their temperature scale. Therefore we estimated the uncertainties by considering the work of \citet{Johnson65}. The standard deviation of the fit in \citet[figure~4]{Johnson65}, approximately $300$\,K, was found by eye. This is adopted as uncertainty for this work. We consider the scale of \citet{CK79} to be a derived scale.


\subsection{\citet{dJN87}}
\label{sect:djn87}
Using a large number of stars with known luminosities and effective temperatures, \citet{dJN87} determined the statistical dependencies of luminosity and effective temperature on spectral type and luminosity class, respectively. Spectral types and luminosity classes are represented by continuous variables. A large number of stellar temperatures was approximated by Chebychev polynomials in the continuous variables. The resulting temperatures are tabulated in their table~5.

For the application to the analysis of T Tauri stars, we extracted the approximated temperatures for dwarfs, subgiants and giants separately from their table~5. The standard deviation of the fit in \citet{dJN87} is ${\Delta}{\log}T_{\mbox{eff}}=0.021$, as given by the authors. This value is used as the intrinsic error of the obtained dwarf, subgiant and giant scales.


\subsection{\citet{HSS94}}
\label{HSS94}
According to the authors, the scale in their table~4 is based on the work of \citet{BB88} and \citet{SK82}. However, temperatures could not be found in the former. For K7 and M types they are identical to the temperatures in \citet[table~2]{Bessell91} (cf. Sect. \ref{sect:bessell91}). For types earlier than K7, the dwarf scale of \citet{SK82} was adopted with the adopted value at K5 being higher by 50\,K.

Consequently, as \citet{HSS94} do not provide intrinsic errors for their scale, we adopt the errors of the combined dwarf scale of \citet{Bessell79} and \citet{Bessell91} and the dwarf scale of \citet{KH95}, the latter being based on \citet{SK82} (see Sect. \ref{sect:kh95}). The intrinsic error of the scale in \citet{Bessell91}, $\pm290$\,K, is adopted for K7 and M spectral types and $\Delta{\log}T_{\rm eff}=\pm0.015$ for earlier types, according to Sect. \ref{sect:kh95}. We consider the scale of \citet{HSS94} to be a derivation of the scales of \citet{SK82} and \citet{Bessell91}.


\subsection{\citet{KH95}, \citet{SK82}}
\label{sect:kh95}
The effective temperatures in \citet[table~5]{KH95} are based on the temperatures given in \citet[table~3, page~453]{SK82}. The latter was derived as described in the following \citep{SK01}: Direct fundamental data were interferometric measurements of stellar diameters \citep[table~22, page~30 in][table~10]{SK82,deJager80}, the sun and \citet[table~6]{deJager80}. Interpolation in spectral type yielded a relation between spectral type and radii \citep[table~23, page~31]{SK82}. Surface brightnesses were added and also interpolated in spectral type. This result was then compared and corrected with the literature. 
The smoothed result is shown in \citet[table~3, page~453]{SK82}. For accuracy reasons, ($U-B$)$_0$ instead of spectral type was used as parameter for very early-type stars, but then transformed to spectral types. Similarly ($R-I$)$_0$ was applied for late-M type stars. In the range B8-K3 V/III, the error is $\Delta{\log}T_{\rm eff}\approx\pm0.015$. For O3-O6 and M5-M8 the errors are much larger.

Following \citet{Kenyon01}, the effective temperature scale is based on integrated spectral energy distributions which are generated by means of the colours in table~5 of \citet{KH95}. Ultraviolet magnitudes are included to obtain reliable spectral energy distributions. If the derived temperatures are close enough to those of \citet{SK82} and \citet{Straizys92}, an appropriate average applies. In case of discontinuities of the scales of \citet{SK82} and \citet{Straizys92} and simultaneous discrepancy with the temperatures, that originate from the spectral energy distributions, the latter are favoured.

For all spectral types with temperatures given in \citet[table~3, page 453]{SK82}, we actually find the same temperatures in \citet{KH95} except for G8, with the temperature being lower by $50\,$K in \citet{KH95}. 


In their appendix B, \citet{KH95} provide a relative error of 5 to 10\,\% for their entire table~5 affecting colours, effective temperatures and bolometric corrections. However \citet{Kenyon01} recommends to obtain errors from \citet{SK82} and \citet{Straizys92}. The error $\Delta{\log}T_{\rm eff}\pm0.015$ \citep{SK01} is applied in our work. We deem \citet[table~5]{KH95} to be an improvement of the work of \citet{SK82}.


\subsection{\citet{Perrin98}}
\label{sect:perrin98}
\citet{Perrin98} extend the effective temperature scale for giants to spectral types later then M6 using interferometrically determined radii of nine giant stars. The temperature scale is tabulated in their table~5. 

The temperatures for spectral types earlier than M6 are based on the work of \citet{Ridgway80} (cf. Sect. \ref{sect:bessell79}). These are considered to be most consistent with previous effective temperature scales and the \citet{Perrin98} data. Correspondingly, for the earlier spectral types we applied an error of $\pm220\,$K.

\citet{Perrin98} derived temperatures for spectral types later than M6. The temperature for M6 is identical to the effective temperature of EU Del. For M8, the average of the temperatures of \object{SW~Vir} an \object{RX~Boo} is adopted. The value for M7 is found by interpolation. The resulting errors are $\pm79\,$K for M6, $\pm94\,$K for M7 and $\pm42\,K$ for M8. Though those errors are only individual errors of single star and any possible statistical scatter is not accounted for, we adopt those errors for consistency reasons. The scale is based on temperatures of individual stars and can be regarded as a primary scale.


\subsection{\citet{Luhman99,Luhman03}}
\label{sect:luhman99}
\citet{Luhman99} provides in table~2 effective temperature scales for dwarfs, giants and intermediate luminosity classes for spectral types later than M0. The dwarf and intermediate scales were updated by \citet[][table~8]{Luhman03}. 
No errors are given for those temperature scales and were estimated by us as explained below. 


\subsubsection{Dwarf scale}
The dwarf scale in \citet{Luhman99} is based on \citet{LR98} who linearly fitted data from \citet{Leggett96}. Moreover, the scale in \citet{Luhman99} is consistent with modelling results of \citet{Leggett98}. For the assessment of errors we considered the work of \citet{Leggett96}. They derived effective temperatures of low-mass stars by fitting synthetic spectra \citep{AH95} to observed low-resolution spectra. Spectral types come from \citet{Henry94}, \citet{Leggett92} and Boeshaar \& Liebert. The errors of the temperatures are $\pm150\,$K and $\pm250\,$K, respectively \citep[table~7]{Leggett96}.

\cite{Luhman03} improved this temperature scale by adjusting the temperatures to be consistent with the latest temperature estimates for young disk dwarfs \citep{Leggett00,Burgasser02}: -50\,K at M5 and -100\,K at M6-M9 (see their figure~8).

The \cite{Leggett96} data allowed us to calculate the standard deviation of the fit in \citet{LR98}. We consider this value of $\pm80\,$K an overall estimate for the uncertainty of dwarf scale in \cite{Luhman03}.

\subsubsection{Giant scale}
The giant scale in \citet{Luhman99} is adopted from \citet{vanBelle99} for spectral types earlier than M7, from \citet{Perrin98} (see Sect.~\ref{sect:perrin98}) for spectral types M7 and M8, and from \citet{Richichi98} for spectral type M9. \citet{Richichi98} and \citet{vanBelle99} inferred effective temperatures directly by using apparent angular diameters that originate from lunar occultation measurements and interferometric measurements, respectively. 

Van Belle et~al. (1999) provide the standard deviation $270\,$K of a linear fit as an estimation of the uncertainty of their scale. The corresponding errors of the giant scale of \citet{Luhman99} are $270\,$K for spectral types earlier than M7, $\pm100\,$K for M7, $\pm50\,$K for M8 and $\pm150\,$K for M9. We consider this scale a derived scale.

\subsubsection{Intermediate scale}
\label{sect:luhman99i}
\citet{Luhman99} created an intermediate temperature scale from the temperatures and spectral types of the \object{GG~Tau} components. The temperatures were obtained by following the steps of \citet{White99}, comparing the luminosities of the components to evolutionary models in the HR diagram. The assumption of coevality requires the components to be located on the same isochrones. Therefore, their effective temperatures were allowed to vary in order to find a coeval solution (see Sect.~\ref{sect:method} for a discussion). With correct temperatures all components should be placed on the same model isochrone. This isochrone is fixed by \object{GG~Tau} Aa \& Ab with spectral types K7 and M0.5, respectively, as giant temperatures and dwarf temperatures are not very different in this range of spectral types. A further constraint is the total mass of \object{GG~Tau~A} which was determined by \citet{Guilloteau99} from the Keplerian rotation of the circumbinary disk. An evolutionary model is then considered to be consistent with \object{GG~Tau} if \object{GG~Tau~A} fulfils the mass constraint and if the fixed isochrone yields a solution for the temperatures of the low-mass components \object{GG~Tau~B}a \& Bb. 

\citet{Luhman99} found coevality of all GG Tau components when using \citet{BCAH98} isochrones and effective temperatures of $3057\,$K and $2805\,$K for GG Tau Ba and Bb at spectral types M5.5 and M7.5, respectively. The temperature scale was extrapolated from M5.5 to M0 ($3850\,$K). Temperatures for M8 and M9 are chosen to be intermediate between giants and dwarfs and to provide continuity of the intermediate scale.

The intermediate scale of \citet{Luhman99} is only valid for M stars. For young stars with earlier spectral types, an intermediate scale is not deemed to be necessary because they evolve quickly towards the main sequence. Their luminosity class is only for a short time between that of giants and dwarfs, whereas M stars develop much slower and have an intermediate character for a correspondingly longer time span. Our analysis allows to check this assumption as we show in Sect.~\ref{sect:application}.

Using \citet{White99}, we assessed intrinsic errors for the intermediate scale of \citet{Luhman99}. \citet{White99} provide error bars for the effective temperatures of the \object{GG~Tau} components. Those errors are attributed to uncertainties in the spectral types of the components: $\pm0.5\,$ subclasses for Ba and Bb. Direct assessment of the effective temperature error bars in \citet[figure~6]{White99} by eye yielded $\pm80\,$K and $\pm100\,$K, resp. The uncertainties of the evolutionary models should also be considered since the intermediate temperature scale is created by means of evolutionary models. The comparison of the \citet{BCAH98} models for $l/H_{\rm P}=1.9$ and $l/H_{\rm P}=1.0$ should provide a simple upper limit for error estimates. Corresponding to \citet{White99}, coevality of the \object{GG~Tau} components using the model with $l/H_{\rm P}=1.0$ yields effective temperatures of $3160\,$K and $2840\,$K for GG Tau Ba and Bb, respectively. That differs by $110\,$K and $20\,$K, resp. from the values derived with the model for $l/H_{\rm P}=1.9$ ($3050\,$K and $2820\,$K, respectively). This difference is of the same order as the uncertainties derived from the spectral types or even lower. Hence, the errors which originate from the uncertainties of the spectral types alone should provide a rough estimate for the intrinsic errors of the intermediate temperature scale of \citet{Luhman99}. We adopt $\pm80\,$K for spectral types M earlier than M6.5 and $\pm100\,$K later than M6.5.

\citet{Luhman03} improved the intermediate temperatures of \citet{Luhman99} by adjustments at spectral types M7-9, so that the sequences of IC\,348 and Taurus are coeval in the Hertzsprung-Russell diagram when comparing to the evolutionary models of \cite{BCAH98} and \cite{CBAH00}. For this adjustment, the temperatures were reduced by 10\,K at M7 and M8 and by 150\,K at M9.


\subsection{\citet{Tokunaga00}}
The dwarf temperature scale of \citet{Tokunaga00} is an average of scales of other authors and therefore considered to be a derived scale within the frame of this work. For the relevant late spectral types the scale results from averaging the values of \citet{Popper80}, \citet{BV81}, \citet{BV82}, \citet{Blackwell91}, \citet{BG89}, \citet{Bessell91}, \citet{Jones95} and \citet{Leggett96}. An intrinsic error of $\pm100\,$K is provided for the later spectral types. 

\citet{Popper80} uses temperatures of \citet{Hayes78} and fluxes of \citet{Barnes78}. The temperatures in \citet{Hayes78} are based on \citet{Code76} (cf. Sect. \ref{sect:bessell79}) and \citet{Conti73}. \citet{BV81} provides temperatures based on different publications. \citet{Blackwell91} derive temperatures with the infrared flux method and improved H$^-$-opacities. \citet{BG89} use the infrared flux method and synthetic infrared colours to derive temperatures for G- and K-type stars. \citet{BV82} infers temperatures for spectral types A-F applying a temperature sensitive discontinuity of the spectral energy distribution at $1600\,${\AA}. \citet{Jones95} compared observed and synthetic spectra for transitions of water vapour in order to provide temperatures for M-dwarfs. 


\section{Conversion tables}
\label{app:tables}
\begin{table*}[H]
\caption{The temperature conversion scales for dwarfs at spectral types G0-M9. For details on the references see Table~\ref{tab:scales}. Temperatures are given in [K].}
\label{tab:dwarfs}
\centering
\begin{tabular}{c|cccccccc}
\hline
\hline
spectral type     &bessell7991$^1$&ck79$^2$&djn87$^3$&hss94$^4$
&kh95$^5$&luhman03$^6$&tokunaga00$^7$\\
\hline
G0   &6000         &5900 &5940  &6026      &6030 &         &5930       \\
\hline
G1   &             &5830 &      &          &5945 &         &           \\
\hline
G2   &             &5770 &5790  &5860      &5860 &         &5830       \\
\hline
G3   &             &     &      &          &5830 &         &           \\
\hline
G4   &             &     &5640  &          &5800 &         &5740       \\
\hline
G5   &             &5660 &      &5754      &5770 &         &           \\
\hline
G6   &5500         &     &      &          &5700 &         &5620       \\
\hline
G7   &             &     &      &          &5630 &         &           \\
\hline
G8   &             &5450 &5310  &5572      &5520 &         &           \\
\hline
G9   &             &     &      &          &5410 &         &           \\
\hline
K0   &             &5240 &5150  &5248      &5250 &         &5240       \\
\hline
K1   &             &5110 &4990  &5082      &5080 &         &           \\
\hline
K2   &5000         &4950 &      &4898      &4900 &         &5010       \\
\hline
K3   &             &4780 &4690  &4732      &4730 &         &           \\
\hline
K4   &4500         &4580 &4540  &4592      &4590 &         &4560       \\
\hline
K5   &             &4400 &4410  &4395      &4350 &         &4340       \\
\hline
K6   &             &4200 &      &          &4205 &         &           \\
\hline
K7   &4000         &4000 &4150  &3999      &4060 &         &4040       \\
\hline
K7-M0&             &3960 &      &          &     &         &           \\
\hline
K9   &             &     &3940  &          &     &         &           \\
\hline
M0   &3800         &3920 &3840  &3802      &3850 &         &3800       \\
\hline
M0.5 &             &3800 &      &3724      &     &         &           \\
\hline
M1   &3650         &3680 &3660  &3648      &3720 &3680     &3680       \\
\hline
M1.5 &             &3590 &3590  &3573      &     &         &           \\
\hline
M2   &3500         &3500 &3520  &3499      &3580 &3510     &3530       \\
\hline
M2.5 &             &3430 &      &          &     &         &           \\
\hline
M3   &3350         &3360 &3400  &3350      &3470 &3350     &3380       \\
\hline
M3.5 &             &3300 &      &          &     &         &           \\
\hline
M4   &3150         &3230 &3290  &3148      &3370 &3180     &3180       \\
\hline
M4.5 &             &     &      &          &     &         &           \\
\hline
M5   &3000         &3120 &3170  &2999      &3240 &2960     &3030       \\
\hline
M5.5 &2900         &3040 &      &          &     &         &           \\
\hline
M6   &2800         &2960 &3030  &2799      &3050 &2740     &2850       \\
\hline
M6.5 &2700         &     &2950  &          &     &         &           \\
\hline
M7   &2600         &     &2860  &          &     &2620     &           \\
\hline
M7.5 &2450         &     &      &          &     &         &           \\
\hline
M8   &             &     &2670  &          &     &2500     &           \\
\hline
M9   &             &     &2440  &          &     &2300     &           \\
\hline
\end{tabular}
\begin{tabular}{ll}
$^1$bessell7991\,$=$\,combination of \citet{Bessell79} and \citet{Bessell91}\\
$^2$ck79\,$=$\,\citet{CK79}&$^3$djn87\,$=$\,\citet{dJN87}\\
$^4$hss94\,$=$\,\citet{HSS94}&$^5$kh95\,$=$\,\citet{KH95}\\
$^6$luhman03\,$=$\,\citet{Luhman03}&$^7$tokunaga00\,$=$\,\citet{Tokunaga00}
\end{tabular}
\end{table*}

\begin{table*}[H]
\caption{Same as Table~\ref{tab:dwarfs} for the non-dwarf luminosity classes. The reference abbreviations are appended with `g' for a giant scale, `sg' for a subgiant scale and `i' for an intermediate scale. See Tab.~\ref{tab:scales} for further details. Again temperatures are given in [K].}
\label{tab:giants}
\centering
\begin{tabular}{c|cccccc}
\hline
\hline
spectral type&bessell79g$^1$&djn87sg$^2$&djn87g$^3$&perrin98g$^4$ &luhman03i$^5$&luhman99g$^6$\\
\hline
G0   &          &5640   &5470  &          &          &         \\
\hline
G1   &          &       &      &          &          &         \\
\hline
G2   &          &5460   &5300  &          &          &         \\
\hline
G3   &          &       &      &          &          &         \\
\hline
G4   &          &5280   &5130  &          &          &         \\
\hline
G5   &          &       &      &          &          &         \\
\hline
G6   &          &       &      &          &          &         \\
\hline
G7   &5000      &       &      &          &          &         \\
\hline
G8   &          &4940   &4800  &4930      &          &         \\
\hline
G9   &          &       &      &          &          &         \\
\hline
K0   &4750      &4780   &4660  &4790      &          &         \\
\hline
K1   &          &4620   &4510  &4610      &          &         \\
\hline
K2   &4500      &       &      &4450      &          &         \\
\hline
K3   &4250      &4340   &4260  &4270      &          &         \\
\hline
K4   &          &4210   &4150  &4095      &          &         \\
\hline
K5   &4000      &4080   &4050  &3980      &          &         \\
\hline
K6   &          &       &      &          &          &         \\
\hline
K7   &          &3870   &3870  &          &          &         \\
\hline
K7-M0&          &       &      &          &          &         \\
\hline
K9   &          &3700   &3740  &          &          &         \\
\hline
M0   &          &3630   &3690  &3895      &          &         \\
\hline
M0.5 &          &       &      &          &          &         \\
\hline
M1   &          &3510   &3600  &3810      &3705      &3800     \\
\hline
M1.5 &          &3460   &3560  &          &          &         \\
\hline
M2   &3750      &3410   &3540  &3730      &3560      &3700     \\
\hline
M2.5 &          &       &      &          &          &         \\
\hline
M3   &          &3340   &3480  &3640      &3415      &3590     \\
\hline
M3.5 &          &       &      &          &          &         \\
\hline
M4   &          &3280   &3440  &3560      &3270      &3480     \\
\hline
M4.5 &3500      &       &      &          &          &         \\
\hline
M5   &          &3220   &3380  &3420      &3125      &3370     \\
\hline
M5.5 &          &       &      &          &          &         \\
\hline
M6   &3250      &3150   &3330  &3243      &2990      &3250     \\
\hline
M6.5 &          &3110   &3300  &          &          &         \\
\hline
M7   &          &3070   &3270  &3087      &2880      &3100     \\
\hline
M7.5 &          &       &      &          &          &         \\
\hline
M8   &          &2990   &3240  &2806      &2710      &2800     \\
\hline
M9   &          &2920   &3270  &          &2400      &2650     \\
\hline
\end{tabular}
\begin{tabular}{ll}
$^1$bessell79g\,$=$\,\citet{Bessell79}&$^2$djn87sg\,$=$\,\citet{dJN87}\\
$^3$djn87g\,$=$\,\citet{dJN87}&$^4$perrin98g\,$=$\,\citet{Perrin98}\\
$^5$luhman03i\,$=$\,\citet{Luhman03}&$^6$luhman99g\,$=$\,\citet{Luhman99}\\
\end{tabular}
\end{table*}


\end{document}